\documentclass[twocolumn,aps,prb,showpacs,floatfix]{revtex4}

\usepackage{graphicx}%
\usepackage{dcolumn}
\usepackage{bm}%

\begin{document}
\title{Magneto-Optical Properties of Bound Excitons in ZnO }

\author{A.~V.~Rodina,}
 \altaffiliation{permanent address: A. F. Ioffe Physico-Technical
Institute, 194021, St.-Petersburg, Russia.} 
\author{ M.~Strassburg, M.~Dworzak, U.~Haboeck, A.~Hoffmann,}
\affiliation{Institute
for Solid State Physics, TU Berlin, D-10623 Berlin, Germany}
\author{A.~Zeuner, H.~R.~Alves, D.~M.~Hofmann, B.~K.~Meyer}
\affiliation{I.~Physics Institute, Justus-Liebig-University of  
Giessen,
35392 Giessen, Germany}

\date{\today}

\begin{abstract}

We  present  results of magneto-optical  measurements  and
theoretical analysis of shallow bound exciton complexes in bulk
ZnO.   Polarization  and  angular  dependencies   of   
magneto-photoluminescence  spectra at 5 T suggest  that  the  upper
valence  band  has  $\Gamma_7$  symmetry.  Nitrogen  doping  leads   
to the 
formation  of an acceptor center  that  compensates 
shallow  donors. This is confirmed by the observation  of  excitons
bound to ionized donors  in nitrogen doped ZnO. The
strongest  transition in the ZnO:N  ($I_9$  transition)  is
associated  with  a  donor bound exciton. This conclusion is based on  
its  thermalization
behavior    in    temperature-dependent    magneto-transmission
measurements and is supported by  comparison   of
the  thermalization  properties of the $I_9$ and $I_4$ emission lines 
in   temperature-dependent
magneto-photoluminescence   investigations.

\end{abstract}
\pacs{78.20.Ls,78.55.-m,78.40.-q,71.35.Ji}

\maketitle

\section{ Introduction}

    The  near  band  gap photoluminescence (PL) and  transmission (PT) 
spectra  of  bulk ZnO are known to exhibit a large  variety  of
lines  (usually numbered from $I_0$ to $I_{11}$) stemming from 
excitons
bound  to  charged and neutral impurity centers.\cite{1,2,3,4} 
Although
the  origin of these transition lines remained in the  focus  of
intensive research over more than 30 years,\cite{1,2,3,4,5,6,7,8,9} a  
convergent
picture  concerning  the  nature of the  impurities  (donor  or
acceptor character) as well as the symmetry of the valence band
hole  ($\Gamma_7$  or $\Gamma_9$ character) involved in the complexes, 
has  not
yet emerged.

    Bulk  ZnO  is a direct band gap  semiconductor where
 the  valence band maximum is split in a triplet (denoted as $A$, $B$ 
and $C$) by the wurtzite crystal field  and
spin-orbit  interaction.  The symmetry  of  the  upper  valence
subband ($A$ subband) has been the subject of  a controversy 
($\Gamma_9$  or
$\Gamma_7$  character)  for more than 40 
years.\cite{10,11,12,13,14,15,16,17,18,19,20}
  Based  on  the
polarization  properties of the free exciton transitions,  most
of  the  authors  assumed that the symmetry of  the  $A$  valence
subband  in  bulk ZnO is $\Gamma_7$,\cite{1,2,3,10,11,12,14,15,16,17} 
in contradiction
to  most recent studies of the free exciton oscillator
strengths (see Ref. \onlinecite{18}) and  magneto-optical studies of 
the free $A$
exciton   fine  structure  (see Ref.  \onlinecite{19}). They  were  
interpreted   
assuming  that the valence  band  maximum has $\Gamma_9$ symmetry.
More  elaborated  theoretical  analysis  of  the  exciton  fine
structure  made  in Ref. \onlinecite{20}, enabled the explanation  of  
the
magneto-optical data of Ref. \onlinecite{19} assuming $\Gamma_7$ 
symmetry for  the
$A$  valence  subband. A self--consistent set of the  effective  mass
parameters obtained in the first principles calculations  along
with the respective structure of the valence band has been used
in  this  analysis.\cite{20} The calculated value of  the  $\Gamma_7$  
hole
effective  $g$  factor  of  $g_h^\parallel = - 1.3$ in  the  magnetic  
field
parallel  to  the $c$ axis (${\bm B} \parallel c$) is close to the 
value 1.2  assumed  in
Ref. \onlinecite{19} for the $\Gamma_9$ hole but has the opposite 
sign. On the contrary, the
calculated  value of the $\Gamma_9$ hole effective g factor  
$g_h^\parallel \approx 3.0$
 (see Ref. \onlinecite{20})  does not allow to describe the splitting 
observed in Ref.
\onlinecite{19}.

    Magneto-optical studies are a powerful tool to  elucidate
the   origin  of  the different  bound  exciton  transitions.   The
theoretical   basis  for  the  description  of  magneto-optical
properties  of  excitons  bound  to  neutral  or   charged
impurities  in hexagonal semiconductors has been  developed  by
Thomas  and  Hopfield.\cite{21} The identification  of  excitons  
bound  to ionized  impurities  is
enabled  by  the  nonlinear splitting  of transitions in  the  
magnetic  field
perpendicular to the $c$ axis (${\bm B} \perp c$). Such splitting 
governed by the zero field
 spin--exchange  splitting  of  the  corresponding  exciton  state was 
indeed observed in ZnO for $I_2$ and $I_3$ lines.\cite{3,4}   
The  transitions  from  excitons bound to  neutral  donors ($(D^0,X)$ 
complex) can be distinguished from 
  the excitons bound to neutral acceptors ($(A^0,X)$ complex) by the 
thermalization of the absorption or emission components in magnetic 
field  ${\bm B} \perp c$. The corresponding thermalization in 
absorption is caused by the splitting of the ground state ($D^0$ 
state) which was observed for the lines from $I_5$ to  $I_9$ in Ref. 
\onlinecite{3}. An additional proof for the $(D^0,X)$ origin of these 
lines, as well as of the $I_4$ line, was obtained by the observation  
of  two  electron
satellite  (TES) PL lines\cite{5,6,7,8,helde,overview}  that are the 
transitions to the ($n=2$) excited state of the neutral donor. 
However, the authors of  Ref. \onlinecite{1} reported on the thermalization  of   the   magneto-PL components  in ${\bm B} \perp c$ 
for the lines  $I_5$ to$ I_{11}$. The thermalization in PL is caused  
by the splitting of the  excited state of the acceptor bound exciton 
complex. Therefore, an attribution of  these lines $I_5$ to $I_{11}$  
to $(A^0,X)$ complexes was given in Ref. \onlinecite{1}. This is in 
contradiction to the magneto-absorption data of Ref. \onlinecite{3}. 
Moreover, from the thermalization of PL components in magnetic field 
${\bm B} \parallel c$ 
the  authors of Refs. \onlinecite{1}  concluded that the Zeeman 
splitting of the
hole  in the acceptor ground state was  larger than  the
Zeeman  splitting of the electrons. This contradicts strongly to the 
results reported previously in Ref. \onlinecite{2} for the $g$ values 
of the $\Gamma_7$ holes involved in the same transitions.  Thus, the 
nature of the neutral bound exciton transitions, the symmetry of the 
holes and  the reliable hole $g$ values  are up to now not clear.

    In  the present work, we report in detail on  theoretical and
experimental  studies  of  the  magneto-optical  properties  of
shallow  bound  excitons in ZnO. We take into account all possible 
 configurations of bound exciton complexes  involving
 holes of $\Gamma_9$ and  $\Gamma_7$ symmetry with different signs and 
values
of  the  hole g factors. For each configuration we analyze  the
selection  rules  for  the optical transitions and  calculate  the
magnetic  field  dependencies of  the  transition  energies  in
${\bm B} \parallel c$ and ${\bm B} \perp c$ geometry as well  as  
their
dependence on the angle between the magnetic field  and  the  $c$
axis  (angular  dependence).  The 
  analysis of the polarization properties and angular dependencies of 
the magneto-PL transitions in  the  nominally  undoped   and
nitrogen doped ZnO crystals allows us  to reveal the symmetry and the 
$g$ values of the
  valence  band holes.  This enable us further to analyze  the  
thermalization
properties of the magneto-PL and magneto-absorption components in 
magnetic field  perpendicular  and
parallel  to the $c$ axis in order to reveal the nature of the neutral 
impurity centers.

    The  paper  is  organized  as  follows. Section \ref{II} describes   
a
the experimental conditions and the investigated
samples. In Section \ref{III} we report on  the zero  field
PL and PT investigations in ZnO and discuss the Zeeman 
splitting of
the  observed transitions in terms of excitons bound to ionized 
impurity  centers (see Section \ref{IIIA}) and of excitons bound  to
neutral   impurities   centers  (see  Section   \ref{IIIB}).   The
thermalization  properties  of the excitons  bound  to  neutral
impurities are discussed in Sec. \ref{IV}. In
Section \ref{V}  we discuss our results in terms of the literature. 
The main results are summarized in Section \ref{VI}.

\section{\label{II} Experimental setup and samples}

    We  used nominally undoped (as grown from Eagle-Picher, see for 
details Ref. \onlinecite{overview}) and nitrogen doped bulk ZnO 
crystals. A nitrogen  doped
ZnO sample was prepared  from the as grown one by ion implantation 
($2$ MeV, $10^{13}$ ${\rm cm}^{-
2}$)  followed  by a thermal annealing process at  $900^\circ$ C for  
15
minutes (see also Ref. \onlinecite{overview}).

    The  magneto-PL  measurements  were  performed  at  liquid
helium  temperature in a split-coil magneto  cryostat  allowing
the  variation  of both temperature ($2 - 300$  K)  and  magnetic
field  ($0 - 12$ T). Photoluminescence was excited by the $325$  nm
line  of  a HeCd laser. A 450 W XBO lamp was used as excitation
source   for  the  transmission  investigations.  The  spectral
resolution of the detection system was better than $0.15$ meV.

    The  magneto-PL and magneto-PT measurements  were
performed  in Faraday configuration (magnetic field ${\bm B}$  
parallel
to  the  $c$ axis of the crystal and parallel to the ${\bm k}$-vector  
of
 the  detected light) and in Voigt configuration (${\bm B}\perp 
c\parallel {\bm k}$).
Additionally, the angles between ${\bm B}$ and $c$ axises were varied 
from $0$  to
$90$  degrees  at   fixed magnetic field of $5$ T.  The  circular
polarization  of  the light ($\sigma^+$ and $\sigma^-$) in the Faraday  
configuration
was   analyzed  using    $\lambda/4$  plate  and    linear  polarizer.
Additionally,  PL  spectra with ${\bm B} \parallel c$, ${\bm k}\perp   
c$ and  the  electric
field vector of the luminescence light ${\bm E} \parallel  c$ were 
recorded for
the ZnO:N specimen. Thermalization behavior of the Zeeman split
components of the emission and transmission lines was  revealed
by  temperature-dependent measurements in the Faraday and Voigt
configurations.

\section{\label{III} Zeeman behavior of the bound exciton complexes}

\begin{figure*}[hp]
\begin{center}
\includegraphics*[width=7.5cm,height=8cm]{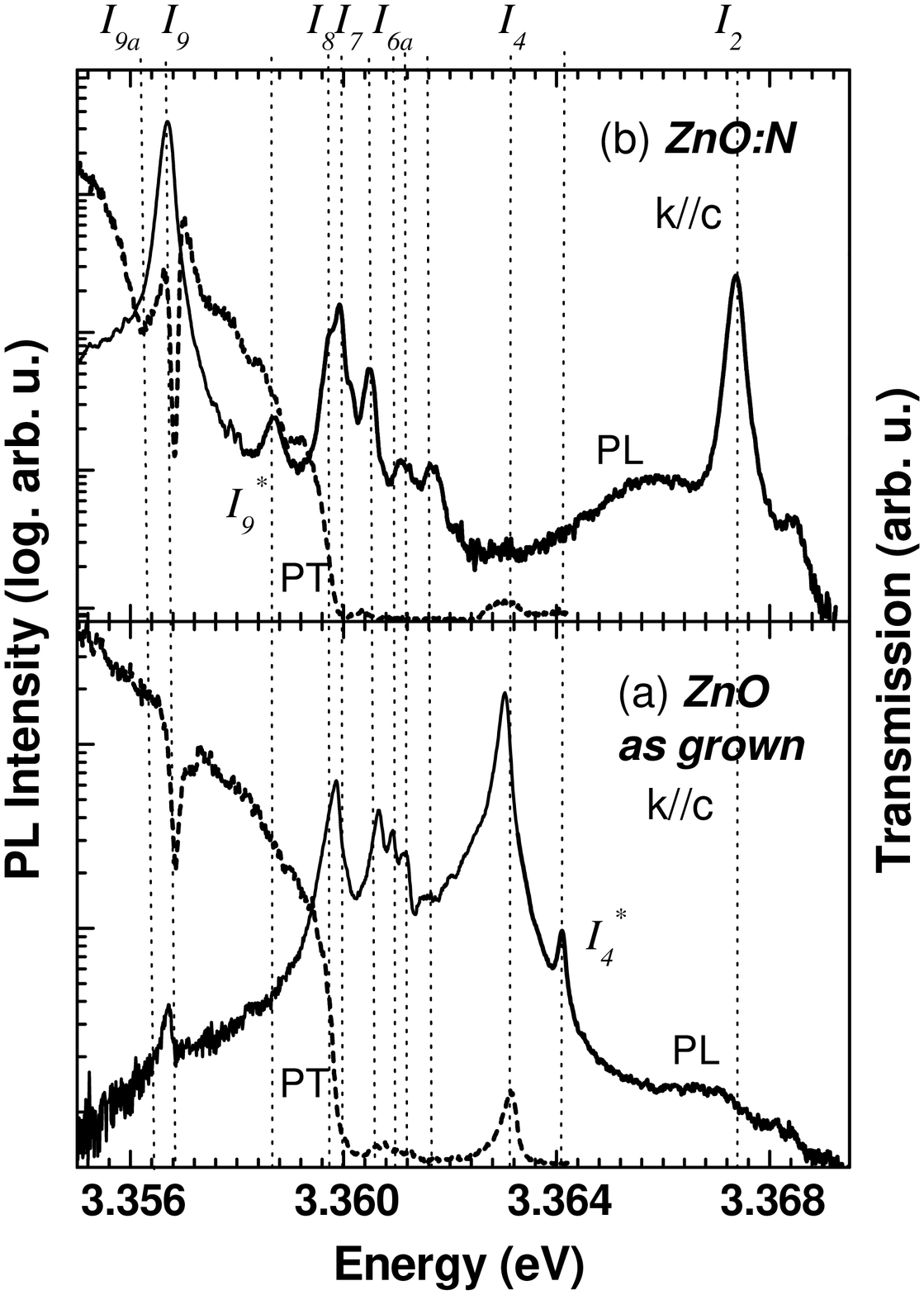}
\end{center}
\caption{\label{Fig1}Photoluminescence  (solid   lines)   
and
       transmission (dashed lines) spectra  of the undoped as grown
       ZnO (a) and nitrogen doped ZnO:N (b) at $4.2$ K}
\end{figure*}

  The
zero  field  PL  (solid line) and PT (dashed  line)
spectra of the as grown (nominally undoped) ZnO and nitrogen  doped
ZnO:N  are  shown  in  Figure \ref{Fig1} (a) and  (b),  respectively.  
The
strongest  lines  in  the  as grown ZnO  (see Fig. \ref{Fig1} 
(a))  are  the  $I_4$
($368.72$   nm   or   $3.3628$   eV)  and the  overlapping   
transitions
attributed  to $I_7$ ($368.90$ nm or $3.3600$ eV) and $I_8$ ($368.92$  
nm
or $3.3598$ eV). Additionally, lines with lower  intensity are
detected  between  $I_4$  and $I_7/I_8$.  The  weak  line $I_9$
($369.27$  nm  or  $3.3567$  eV)  is  observed  both  in   PL   and
PT  spectra. Following the Refs. \onlinecite{helde,overview}, we attribute 
the    $I_4$  transition 
as an   exciton   bound  to a neutral  shallow  H  donor 
that can be easily removed by an annealing 
process.\cite{helde,overview}  Indeed, the $I_4$ PL  line is absent in 
ion implanted ZnO:N  (see Fig. \ref{Fig1} (b)). This is because of the 
post  annealing processes at $900^\circ$ C included in the doping 
procedure.   The  PL spectrum of ZnO:N  is dominated by two lines: a  
new
line  $I_2$  ($368.07$ nm or $3.3676$ eV) and the $I_9$ transition. 
The latter
 considerably gains in intensity  in PL as well as in PT spectra  
compared to the undoped ZnO and becomes  the strongest transition in 
ZnO:N. An  additional transition (labeled $I_{9a}$) on the low-energy 
tail of
the  $I_9$ line in ZnO:N can be seen in the PT spectra that was not 
resolved by the PL measurements and was not observed in the  undoped 
ZnO crystals.
 The wavelengths  of
the  lines  (given  here in air with an accuracy  of  $0.02$  nm)
coincide with the wavelengths and labeling of the lines used in
Refs. \onlinecite{1,2,3,4}.

\begin{figure*}[hp]
\begin{center}
\includegraphics*[width=8cm,height=12cm,angle=270]{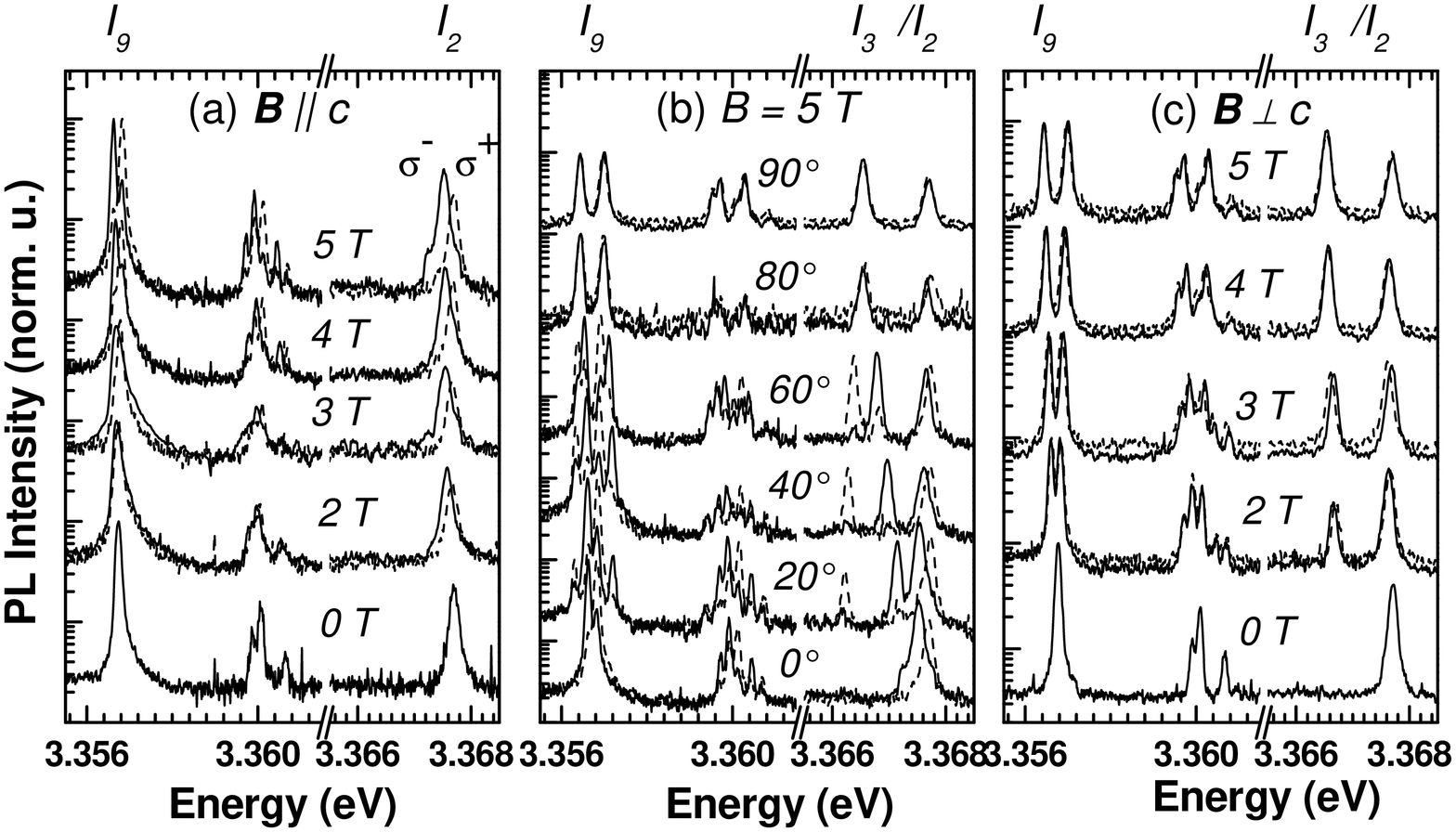}
\end{center}
\caption{\label{Fig2}Photoluminescence  spectra  of  ZnO:N  at   $4.2$ K   
for
       different magnetic fields in (a)  Faraday configuration (${\bm 
B}\parallel  c$
       and ${\bm k} \parallel  c$), (c) Voigt configuration (${\bm 
B}\perp   c$
       and ${\bm k} \parallel  c$),  and for the different angles 
$\Theta$ between $c$ axis and magnetic field direction ${\bm B}$ at
       $5$ T. The spectra measured for right polarized light $\sigma^+$ and left polarized light $\sigma^-$ are shown by dashed and solid lines, respectively.}
\end{figure*}

    In  Figure \ref{Fig2} the magneto-PL spectra 
from   $I_2$  to  $I_9$ for the   ZnO:N    are  shown  in  Faraday
configuration  (a),   Voigt configuration  (c)  and  for  the
different  angles between the magnetic field direction  and  the  $c$
axis at $B=5$ T (b). The most important feature of the spectra in 
Figures \ref{Fig2}  (b) and (c)   is a rising new transition  line at 
an energy about 1 meV below
$I_2$. This $I_3$ transition ($368.19$ nm or $3.3666$ eV in zero 
field), that is  forbidden in ${\bm B} \parallel c$, ${\bm E} \perp c$ 
geometry, has been also observed in parallel magnetic field with the
light  polarized  along the $c$ axis (${\bm E}\parallel c$  
configuration). Such behavior allows us  unambiguously to identify the 
$I_2$ and $I_3$ transition as originating from  excitons bound to an 
ionized impurity  (see a detailed discussion in the Subsection A), 
while the other transitions that show a linear Zeeman splitting in 
${\bm B} \perp c$ will be assigned to excitons bound to neutral 
impurities (see Subsection B).

    \subsection{\label{IIIA}Exciton bound to an ionized impurity}
    
\begin{figure*}[hp]
\begin{center}
\includegraphics*[width=7.5cm,height=8 cm]{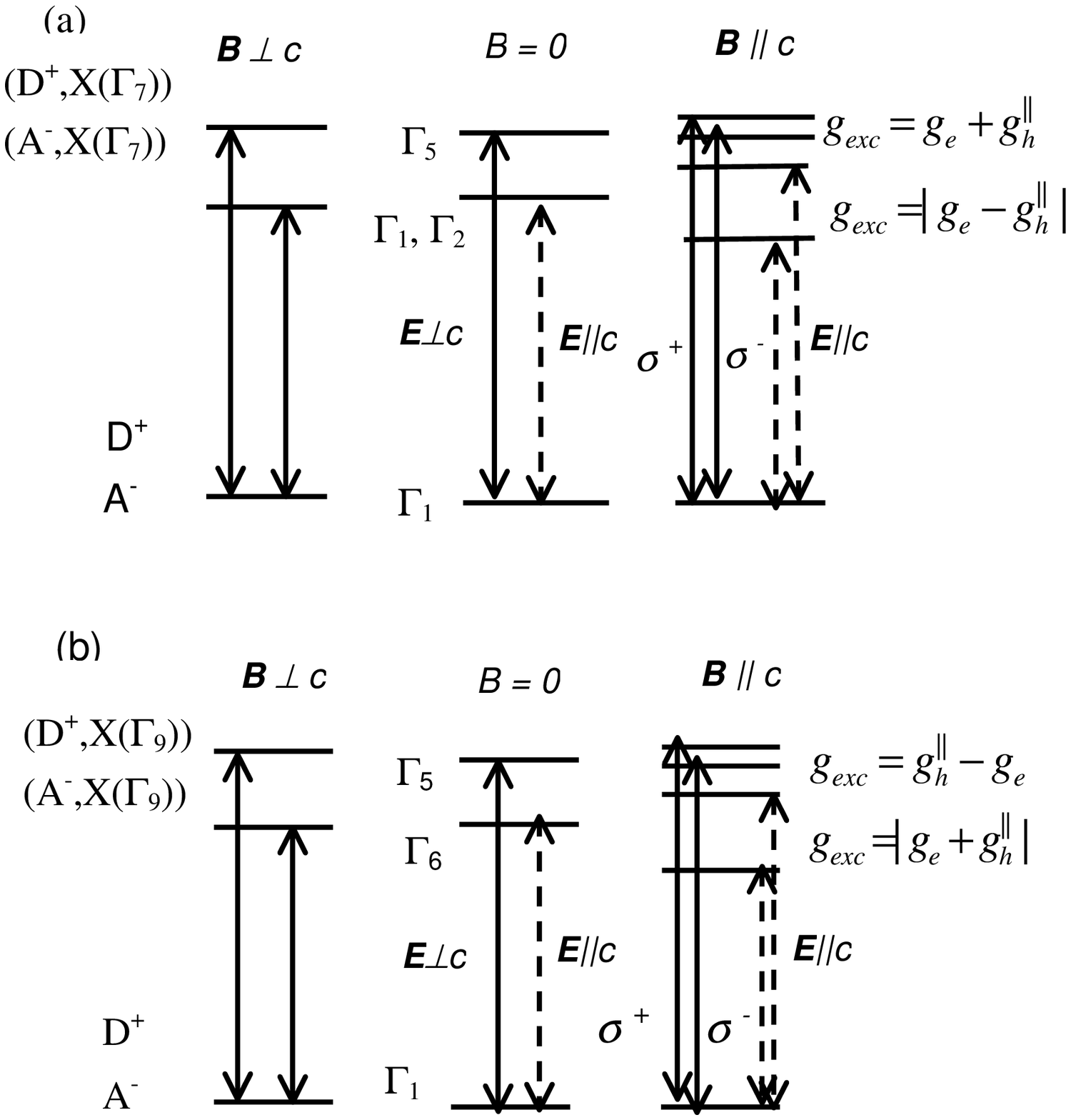}
\end{center}
\caption{\label{Fig3} Level schemes  of ionized bound exciton transitions. Case(a): an exciton involving
       a hole of  $\Gamma_7$ symmetry, and case (b): an exciton 
involving
       a hole of  $\Gamma_9$ symmetry. Note that in (b) the 
transitions with ${\bm E}\parallel c$ are first forbidden but might be 
observed due to high--order perturbations.}
\end{figure*}

\begin{figure*}[hp]
\begin{center}
\includegraphics*[width=5.5cm,height=7.5cm,angle=270]{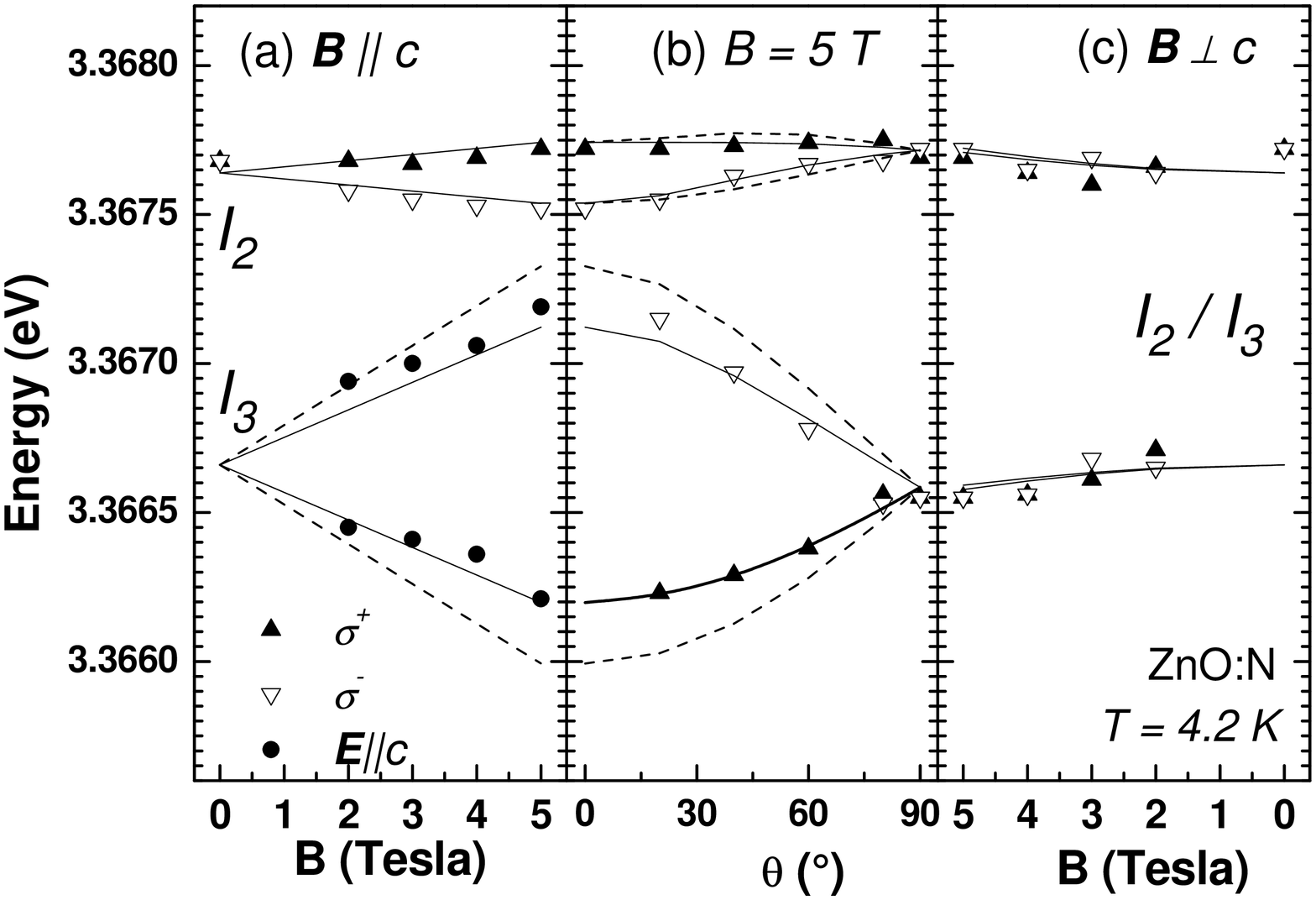}
\end{center}
\caption{\label{Fig4}Zeeman splitting of the bound exciton line $I_2$   
for ${\bm B}
       \parallel c$  (a), ${\bm B}\perp   c$ (c) and its angular 
dependence for $B=5$ T (b).
       Symbols  are experimental data, lines are fits  for  the
       transitions allowed with circular polarized light ($\sigma^+$ 
or $\sigma^-$ ) and
       linear polarized light (${\bm E}\parallel c$), respectively. 
 The hole
       is assumed  to be of  $\Gamma_7$ symmetry with $|g_h^\parallel|<g_e$ (solid lines ) and of 
$\Gamma_9$
       symmetry with $|g_h^\parallel|>g_e$ (dashed lines).}
\end{figure*}

    Let us consider in details  the magnetic field and angular  
dependencies    of  the
emission lines $I_2$ and $I_3$ generated by   excitons bound to an 
ionized impurity.  These dependencies allow  us  to
determine  the  symmetry and the $g$ values of the valence band hole  
involved in the corresponding bound  excitons. 
In Figure \ref{Fig3} we sketch  all possible transitions  with 
selection rules  corresponding to our observations for $I_2$/$I_3$ 
lines in Fig. \ref{Fig2}. We consider the holes from the $\Gamma_7$ 
(case (a)) and $\Gamma_9$ (case (b)) valence bands. The  zero  field 
exchange splitting between  the $\Gamma_1$ and $\Gamma_2$ exciton 
states is neglected in Fig. \ref{Fig3} (a) as it is known to  be small  
in ZnO.\cite{20} 
 The
  magnetic  field  ${\bm B}\parallel c$ linearly splits  the  $I_2$  transition
(allowed  in  ${\bm E}\perp c$ configuration) as well as the $I_3$ 
transition  (allowed  in
${\bm E}\parallel c$ configuration). The
upper-energy component of the $I_2$ line is active  for 
the  right-circular polarized light $\sigma^+$ (see dashed line in 
Fig. \ref{Fig2} (a)). The  perpendicular magnetic field (${\bm B}\perp 
c$)  
mixes the exchange--split exciton states: $\Gamma_5$ state with 
$\Gamma_1/\Gamma_2$ states in the case(a) in Fig. \ref{Fig3} or with 
$\Gamma_6$ state in the case (b). 
The resulting high- and low-energy components of $I_2/I_3$ remain  
nearly
unsplit in ${\bm B} \perp c$ and their energy dependencies on the 
magnetic field are mostly determined by the zero field splitting. Four 
transition components can be distinguished for an
arbitrary angle between the direction of the magnetic field and
the  $c$  axis.  The magnetic field dependencies of the  transition
energies in Faraday and Voigt configurations and  their angular 
dependence at $B=5$ T are  shown  in
Figure \ref{Fig4}.

    The  linear  splitting of the $I_2$ line in ${\bm B} \parallel c$ 
can be described as $\mu_B B  g_{exc}$  with the exciton effective g-
factor
$g_{exc}=0.71$, where $\mu_B$  is the (positive) value  of the Bohr 
magneton.   In
the  case (a), where a hole of $\Gamma_7$ symmetry is involved, we 
have $g_{exc}=g_e+g_h^\parallel$. Here  $g_e \approx 1.95$ is the 
electron (isotropic)  effective  $g$
factor  and  $g_h^\parallel$   is the hole effective $g$ factor  in  
magnetic  field parallel to the $c$  axis.\cite{note1}  In the case 
(b) we have $g_{exc}=g_h^\parallel-g_e$ for a hole of $\Gamma_9$ 
symmetry. Since  $g_{exc} >0$,  we may assume  a
negative effective $g$ factor $g_h^\parallel < 0$, $|g_h^\parallel| < 
g_e$  for the hole of $\Gamma_7$ symmetry   or 
 $g_h^\parallel >  g_e$
for  the  hole of $\Gamma_9$ symmetry. We use the
theoretical expressions obtained in Ref. \onlinecite{21} for the  
angular  dependence  of the
transition energies in Fig. \ref{Fig4}.  From fitting the angular 
dependence and  the
  Zeeman splitting of the $I_3$ line in ${\bm B} \parallel c$  
 it follows that  $|g_h^\parallel | < g_e$. For comparison, the 
dependencies   calculated with  $|g_h^\parallel | < g_e$  and with   
$|g_h^\parallel |  >  g_e$  are shown in  Fig. \ref{Fig4} by solid 
curves  and dashed curves respectively. Since $|g_h^\parallel | < 
g_e$, we see that the 
observed transitions   involve  a  hole of $\Gamma_7$ symmetry as 
shown in Figure \ref{Fig3} (a). This demonstrates that the hole in the 
lowest exciton state and  thus  the upper  $A$
valence  subband in ZnO also have $\Gamma_7$ symmetry. 

    The  values  of the electron and hole effective  g  factors
obtained from the  fitting procedure ( shown by solid curves) in Fig. \ref{Fig4} 
are:  $g_e=1.95$,
$g_h^\parallel   =  -1.24$  and  $g_h^\perp =0.1$.  The value  of  the 
zero-field spin-exchange
splitting  between  the  $\Gamma_5$ exciton state  ($I_2$  transition
allowed  with  the  circular polarized  light)  and  the 
$\Gamma_1/\Gamma_2$ exciton states  
  ($I_3$ transition  allowed with ${\bm E} \parallel c$) is $0.98$  
meV  in  a  good
agreement with Refs. \onlinecite{4,20,24}. The zero field splitting 
between $\Gamma_1$ and $\Gamma_2$ exciton states was neglected. The 
Zeeman splitting of the $I_3$ line in ${\bm B} \parallel c$, however, 
could be fitted more accurately if one assumed the splitting between 
the $\Gamma_1$ and $\Gamma_2$ states to be about $0.1-0.2$ meV. A 
limited spectral resolution in the  ${\bm k} \perp c$, ${\bm E} 
\parallel c$ configuration did not allow us to evaluate this splitting 
more precisely.

The stability of the $(D^+,X)$ or $(A^-,X)$ complex in semiconductors 
depends strongly on the ratio of electron and hole effective 
masses.\cite{lampert} In the case of ZnO, the single band isotropic 
mass approximation predicts the existence of the excitons bound to 
ionized donors.\cite{24} Next, the presence of acceptors in the ZnO:N  
may
induce  the  ionization of the shallow donors\cite{26} as in the case 
of Li and Na doped ZnO.\cite{4} 
Therefore, we attribute the $I_2$ and $I_3$  lines  in  the doped 
ZnO:N  to transitions of excitons bound to  respective
ionized donors ($(D^+,X_A(\Gamma_7))$ complex).

\subsection{\label{IIIB}Excitons bound to neutral impurities}
    
\begin{figure*}[hp]
\begin{center}
\includegraphics*[width=7.5cm,height=9cm]{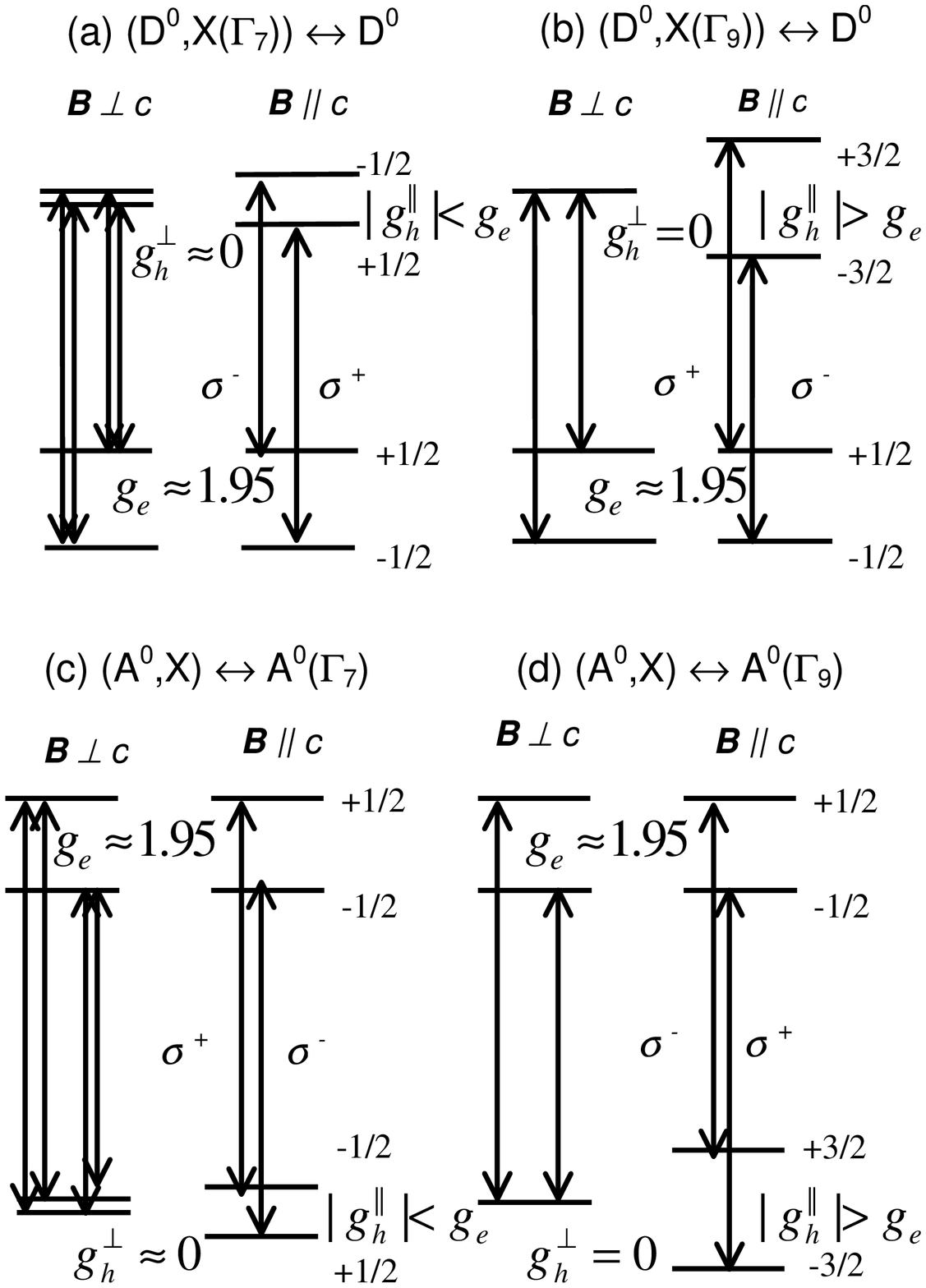}
\end{center}
\caption{\label{Fig5} Level schemes  of neutral bound exciton transitions.  Cases (a) and (b): donor
       bound excitons involving a hole of $\Gamma_7$ symmetry and 
of $\Gamma_9$
       symmetry, respectively. Cases (c) and (d): acceptor bound
       excitons involving a hole of  $\Gamma_7$ symmetry and 
of $\Gamma_9$
       symmetry, respectively.}
\end{figure*}

    The  behavior of the emission lines from $I_4$ to $I_9$  
in ${\bm B} \perp c$ is different from that of the
$I_2/I_3$  lines. Their linear Zeeman splitting indicates 
transitions originating 
from  excitons bound to neutral impurities. In Figure \ref{Fig5} we 
sketch all possible transitions for such complexes  corresponding to 
our observations in Fig. \ref{Fig2}.
  In the case of a donor bound exciton complex
(Fig.  \ref{Fig5}  (a)  and  (b)),  the  spins  of  two  electrons   
are
anti-parallel and the Zeeman splitting of the excited  $(D^0,X)$  
state  is
determined by the anisotropic hole effective $g$ factor ($g_h$), while 
the splitting of the ground $D^0$ state is given by $g_e$.  In the  
case
of  an acceptor bound exciton complex (Fig. \ref{Fig5} (c) and (d)), 
two  holes  have the same  symmetry  and
anti-parallel spins. Hence, the splitting of the excited $(A^0,X)$  
state  is determined  by $g_e$, while the ground $A^0$ state splits 
according to $g_h$. For all transitions from $I_4$ to $I_9$   the 
observed   Zeeman
splitting  in ${\bm B} \parallel c$  is less than the Zeeman splitting 
of the electrons. Furthermore, the upper Zeeman split 
energy  components  of  all transitions are  active in $\sigma^+$ 
polarization. Both of these facts can be explained  in  the
framework  of  our  conclusion that the  ground  state
hole has  $\Gamma_7$ symmetry with  $g_h^\parallel  < 0$,
$ |g_h^\parallel | < g_e$. The
corresponding  ordering of the hole sublevels in  parallel  and
perpendicular magnetic fields and optically active  transitions
in  ${\bm E}\perp   c$ configuration are shown in Fig. \ref{Fig5} (a) 
and (c).
 For  the  sake  of
completeness, we consider also  the $\Gamma_9$ hole with 
 $g_h^\parallel  >  g_e$
(see  Fig. \ref{Fig5} (b) and (d)). The respective transitions 
involving  holes  from  the $B$ valence subband may  be   observed  in  
PL
spectra  at  higher temperatures\cite{helde,overview} or in the 
transmission spectra.

\begin{figure*}[hp]
\begin{center}
\includegraphics*[width=5.5cm,height=7.5cm,angle=270]{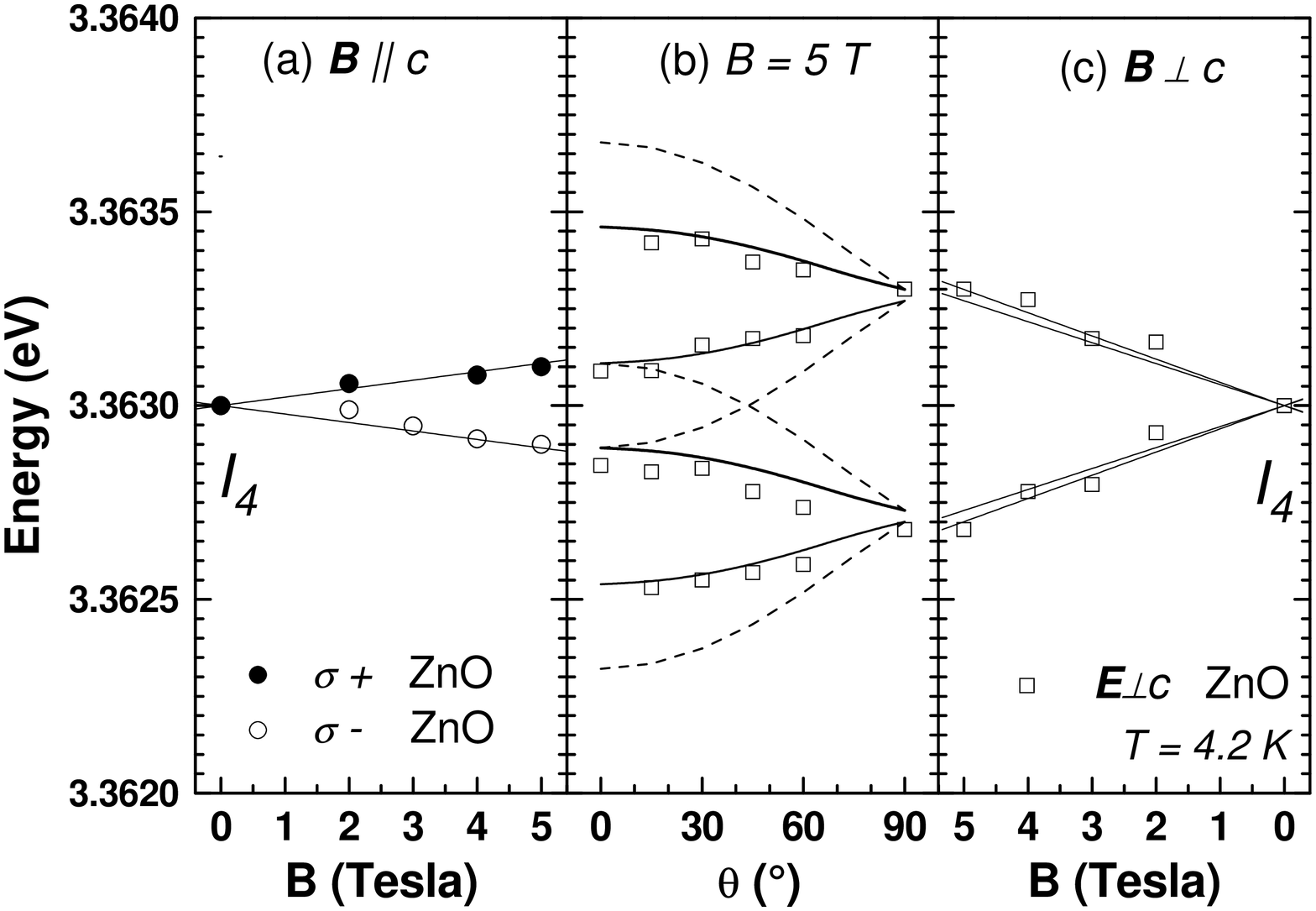}
\end{center}
\caption{\label{Fig6} Zeeman splitting of the bound exciton line $I_4$  
for ${\bm  B}
       \parallel  c $ (a) ,${\bm B}\perp   c$ (c) and its angular 
dependence for $B=5$ T (b).
       Symbols are experimental data and lines are fits for the
       transitions allowed for circular polarized light ($\sigma^+$ 
or $\sigma^-$ ).
        The hole
       is assumed  to be of  $\Gamma_7$ symmetry with $|g_h^\parallel|<g_e$ (solid lines ) and of 
$\Gamma_9$
       symmetry with $|g_h^\parallel|>g_e$ (dashed lines).}
\end{figure*}

\begin{figure*}[hp]
\begin{center}
\vskip 1cm
\includegraphics*[width=5.5cm,height=7.5cm,angle=270]{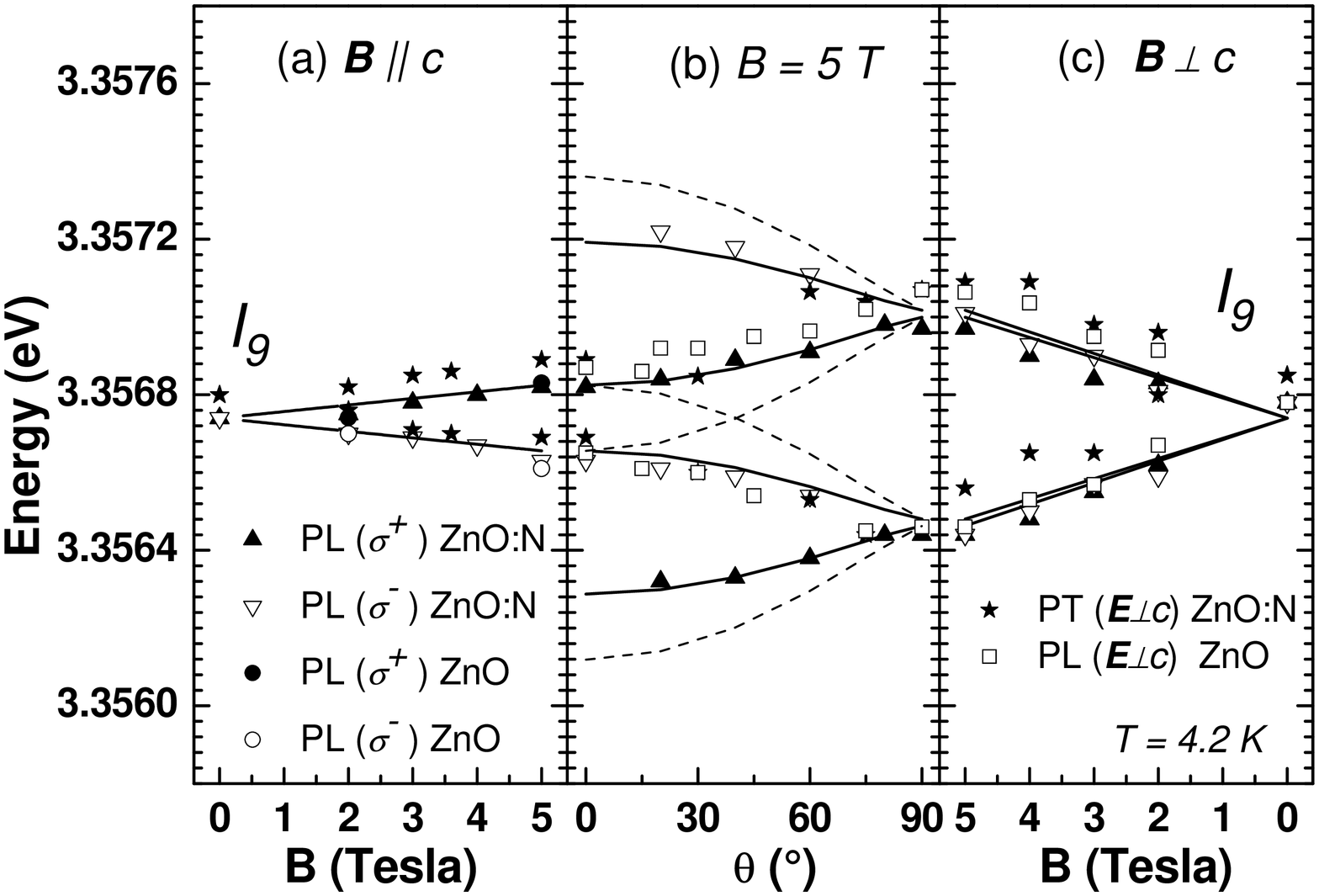}
\end{center}
\caption{\label{Fig7}Zeeman splitting of the bound exciton line $I_9$  
for ${\bm  B} \parallel  c $ (a), ${\bm B}\perp   c$ (c) and its angular 
dependence for $B=5$ T (b).
       Symbols are experimental data and lines are fits for the
       transitions allowed for circular polarized light ($\sigma^+$ 
or $\sigma^-$ ). The hole
       is assumed  to be of  $\Gamma_7$ symmetry with $|g_h^\parallel|<g_e$ (solid lines ) and of 
$\Gamma_9$
       symmetry with $|g_h^\parallel|>g_e$ (dashed lines).}
\end{figure*}

    For a further insight, we  discuss  the emission lines $I_9$ and 
$I_4$ in detail. These lines represent the
  dominant recombination in   ZnO:N  and  as  grown  ZnO  bulk 
crystals,
respectively, and are spectrally resolved   from the other observed 
emission lines.  The
Zeeman  behavior of these two lines is very similar. The transitions
allowed  for  ${\bm E} \perp c$  split linearly as function of 
magnetic  fields
parallel  and  perpendicular to the $c$ axis. Four
transition  components can be distinguished  for  an  arbitrary
angle  between the direction of the magnetic field  and  the  $c$
axis.  The $B$-field dependencies up  to  $5$ T  of the $I_4$ 
transition energies in ${\bm B} \parallel c$ and ${\bm B} \perp c$ as 
well as the angular dependence  at $B = 5$ T are  shown
in  Figure \ref{Fig6}. The corresponding $B$-field and angular 
dependencies of  the  $I_9$
line   in   ZnO:N  are shown in 
Figure \ref{Fig7}.  Additionally, we also show the results of the PT 
measurements for ZnO:N and of the  PL measurements for as  grown
ZnO in  Fig. \ref{Fig7}.

    According to Ref. \onlinecite{21}, the magnetic field dependence  
of  the
transition energies and the absolute value of the hole anisotropic $g$ 
factor
are  given   by  $\pm 1/2 \mu_B B (g_e \pm g_h) \quad$  and
$g_h=\sqrt{|g_h^\parallel|^2 \cos^2\Theta + |g_h^\perp|^2 
\sin^2\Theta}$,
respectively, where   $\Theta$  is the angle between  the  $c$  axis  
and  the
direction  of  the  magnetic  field.  
The analysis of the angular dependence  shows that an excellent 
agreement between experimental data and theoretical modeling is 
possible if and only if we assume that the hole $g$ factor is  smaller 
than  electron's g factor: $|g_h^\parallel| < g_e$. This conclusion 
follows from a comparison  of the angular dependencies 
  calculated  with
$|g_h^\parallel | < g_e$ (solid curves) and with  $|g_h^\parallel | > 
g_e$ (dashed curves) in
Figures \ref{Fig6} and \ref{Fig7}. The values of the electron and hole 
effective
$g$  factors used for the fitting (solid curves) are  $g_e=1.97$, 
$g_h^\parallel  = -1.21$  and
$g_h^\perp =0.1$ for the $I_4$  
and $g_e=1.86$, $g_h^\parallel   = -1.27$ and $g_h^\perp =0.06$ for 
the $I_9$ line. We note that only the data for $I_9$ PL line in ZnO:N 
are used for the fitting procedure in Fig. \ref{Fig7}.  Since 
$g_h^\perp $ is small, it is neither possible  to resolve a respective 
additional splitting of the transitions in ${\bm B} \perp c$ nor to 
observe the nonlinear behavior of the dependence of the transition 
energies on $\cos \Theta$ indicating the holes of the $\Gamma_7$ 
symmetry.\cite{3} Nevertheless, the $\Gamma_7$  symmetry of the holes 
is unambiguously established by the analysis presented above. 

Now, we have found on one hand that the Zeeman behavior of the $I_9$ 
lines in ZnO and ZnO:N  is identical (see Fig. \ref{Fig7}) and very 
similar to the  Zeeman  behavior  of the $I_4$ line in as grown
ZnO. The values of $g_h^\parallel$ derived for $I_4$ and $I_9$ are 
very close to the $g_h^\parallel=-1.24$ obtained in the Subsection A 
for the hole involved in the exciton bound to ionized donor and  to 
the $g$ factor of the hole in $1S$ free exciton 
state.\cite{15,16,17,20} On the other hand, one may expect the  
$g_h^\parallel$ values of the holes involved in the  acceptor bound 
exciton transitions to differ significantly from the $g$ values of the  
holes involved into excitons bound to ionized or neutral donors as it 
was found in CdS.\cite{21} Indeed, the expected value of the 
$\Gamma_7$ hole $g$ factor in a shallow acceptor ground state in ZnO 
is about $-0.75 \pm 0.05$.\cite{note2}
 This makes us to assume that both $I_4$ and $I_9$ transitions are 
rather to be assigned to the 
$(D^0,X_A(\Gamma_7))$   than to the  $(A^0(\Gamma_7),X_A (\Gamma_7))$ 
complexes.  
To  support this assumption, we analyze the thermalization
properties of the  $I_4$ and $I_9$ lines in the next Section.

      We  have  observed
very similar Zeeman behavior for  the  less intensive  lines between 
$I_4$ and $I_9$ with  the
values  of  the  electron and hole effective g factors  in  the
range  of  $g_e=1.92\pm  0.06$, $g_h^\parallel  = -1.24 \pm  0.04$
  and $g_h^\perp =0.08\pm 0.04$.
The  Zeeman splitting of the hole states is found to be  always
smaller  than the splitting of the electron states. From the 
polarization properties it follows then that all observed PL 
transitions involve the $\Gamma_7$ holes from the $A$ valence subband. 
For the hole of $\Gamma_9$ symmetry form the $B$ valence subband the
splitting  in  the parallel magnetic field is predicted  to  be
larger   than   those  of  the  electron.\cite{20}  This would lead to 
the crossing of the transition energies at an angle around $40^\circ$  
as shown by dashed lines in Figs. \ref{Fig6} and \ref{Fig7}. Since we 
do not observed such a crossing, we can not associate any transition 
with the holes of $\Gamma_9$ symmetry. 
We conclude therefore, that the excited states of the complexes 
involving  holes from the  $B$  valence
subband do not make major contribution to the low temperature PL.

\section{\label{IV} Thermalization  properties of excitons  bound  to  
neutral
impurities}

The knowledge of the $\Gamma_7$ hole effective $g$ factors obtained in 
the previous Section  allows us to predict the respective 
thermalization
properties  of  the  Zeeman split components in magnetic fields 
parallel and perpendicular to the $c$ axis (see Figure \ref{Fig5} (a) 
and (c)). We recall, that   the thermalization behavior of the 
absorption components depends on the splitting of the ground ($D^0$ or 
$A^0$) state of the respective complex. In a contrary to that, the 
thermalization of the PL components is caused by the splitting of the 
complex excited state, $(D^0,X)$ or  $(A^0,X)$ respectively. Since 
$g_h^\perp$ of the $\Gamma_7$ hole is small, a  distinct  allocation 
of the observed transitions  either
to  a neutral acceptor or to a neutral donor exciton complex in ZnO, 
similar to the situation in CdS,\cite{21}  can be
facilitated  by thermalization properties in  ${\bm B} \perp c$.

\subsubsection*{Thermalization properties in transmission}

According to Fig. \ref{Fig5} (c), the intensities
of the $A^0(\Gamma_7) \longrightarrow (A^0(\Gamma_7),X(\Gamma_7))$ 
transitions
are  expected  to  be  equal in ${\bm B} \perp c$. In contrast, the 
intensity of the  low-energy  component of the  $D^0 \longrightarrow 
(D^0,X(\Gamma_7))$ transitions in ${\bm B} \perp c$ should  be less 
than that of the high-energy component at low temperatures and should 
increase with  temperature. The opposite behavior for the 
$A^0(\Gamma_7) \longrightarrow (A^0(\Gamma_7),X(\Gamma_7))$  (case (c) 
in Fig. \ref{Fig5})  and $D^0 \longrightarrow (D^0,X(\Gamma_7))$ (case 
(a) in Fig. \ref{Fig5}) transitions is expected in ${\bm B} \parallel 
c$ too. Since  $|g_h^\parallel |<g_e$, 
  the  intensity of the  low-energy  component at low temperatures in 
${\bm B} \perp c$ 
should  be less or more than that of the high-energy component in the 
cases (a) or (c) respectively. The intensity of the
low-energy component (high-energy component) should increase with 
temperature in the case of donor bound exciton (acceptor bound 
exciton) complexes.

    Temperature-dependent transmission spectra  were  recorded
for magnetic fields up to  $5$ T. The absorption of the $I_9$ line
was revealed in both samples. In the ZnO:N the   additional
absorption  peak $I_{9a}$ appeared. However, we  focus  here  only on  
the  main  $I_9$ absorption peak in ZnO:N. The magnetic field and 
angular dependencies  of
this absorption peak derived from PT spectra
(see  Fig. \ref{Fig7}) confirm that it has the same origin as the 
$I_9$ PL line. Unfortunately, we were not able to detect the 
absorption peak corresponding to the PL $I_4$ line in as grown ZnO. 
Instead, we observed an increase of the detected light intensity at 
the respective energy (see Fig. \ref{Fig1} (a))  probably caused by 
the strong emission processes. 

\begin{figure*}[hp]
\begin{center}
\vskip 1cm
\includegraphics*[width=5.5cm,height=7.5cm,angle=270]{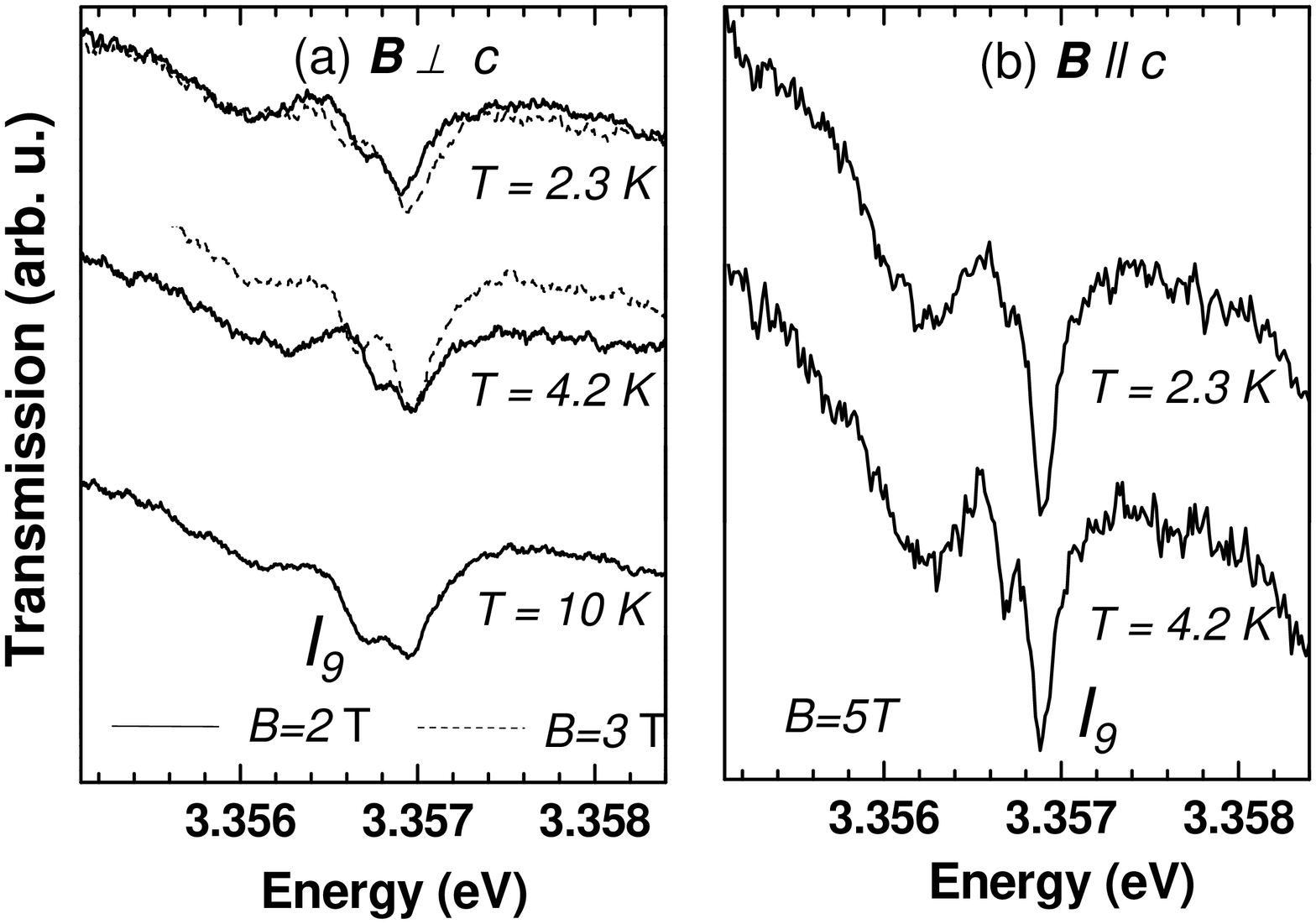}
\end{center}
\caption{\label{Fig8} Temperature-dependent transmission spectra 
of  the bound exciton line 
 $I_9$ in  ZnO:N for magnetic field $B = 2$ T (solid curve) and 
$B=3$ T (dashed curve) perpendicular to the $c$
       axis  of  the crystal (a) and magnetic field  $B  =  5$  T
       parallel to the $c$ axis of the crystal (b). }
\end{figure*}

    The   thermalization  properties  of  the  absorption line $I_9$ 
for ${\bm B} \perp c$ at 
 $2$ T (solid curves) and at $3$ T (dashed curves),  and for ${\bm B} 
\parallel c$ at  $5$ T  are  depicted  in
Figure  \ref{Fig8}  (a)  and (b), respectively. The observed  
temperature
behavior of the low-energy and high-energy absorption components of 
the $I_9$ for ${\bm B} \perp c$ and ${\bm B} \parallel c$ is
in  complete agreement with the expected behavior of the $D^0 
\longrightarrow (D^0,X(\Gamma_7))$
transitions as described above. The intensity ratio of the low-energy 
and the high-energy Zeeman split components decreases with increasing 
magnetic fields (due to the respective increase of the Zeeman 
splitting of the initial $D^0$ state) and increases with increasing 
temperatures.  Therefore, we attribute   the $I_9$ line as a  donor 
bound exciton complex in an excellent agreement with magneto-
absorption data  of Ref. \onlinecite{3}.

\subsubsection*{Thermalization properties in emission}

    Although the $I_9$ transition in nominally undoped ZnO  single
crystals  was assigned to a donor  bound  exciton
complex  in   Ref. \onlinecite{3}), in more recent  publication of 
Ref. \onlinecite{1}
this line was attributed to an acceptor bound exciton. 
 The latter conclusion  was based on the thermalization behavior of 
the PL components in ${\bm B} \parallel c$.\cite{1} Indeed, in  
PL measurements the thermalization properties are determined by the 
splitting of the excited state of the complex. Therefore, the 
intensities
of the $(D^0,X(\Gamma_7)) \longrightarrow D^0$ transitions (see Fig. 
\ref{Fig5} (a))
are  expected  to  be  equal in ${\bm B} \perp c$, while  the 
intensity of the  low-energy  component of the  $ 
(A^0(\Gamma_7),X(\Gamma_7)) \longrightarrow  A^0(\Gamma_7)$ 
transitions (see Fig. \ref{Fig5} (c)) in ${\bm B} \perp c$ should  be 
less than that of the high-energy component at low temperatures and 
should increase with rising temperature. However, since 
$|g_h^\parallel|<g_e$, the same  or the  opposite intensity relation 
should be also observed in ${\bm B} \parallel c$ in the case of $ 
(A^0(\Gamma_7),X(\Gamma_7)) \longrightarrow  A^0(\Gamma_7)$  or 
$(D^0,X(\Gamma_7)) \longrightarrow D^0$ transition, respectively.  To 
verify the self--consistence of the PL results, we also analyze  the 
thermalization properties of the $I_9$ and $I_4$ PL lines for ${\bm B} 
\perp c$ and ${\bm B} \parallel c$. 
 
\begin{figure*}[hp]
\begin{center}
\vskip 1cm
\includegraphics*[width=5.5cm,height=7.5cm,angle=270]{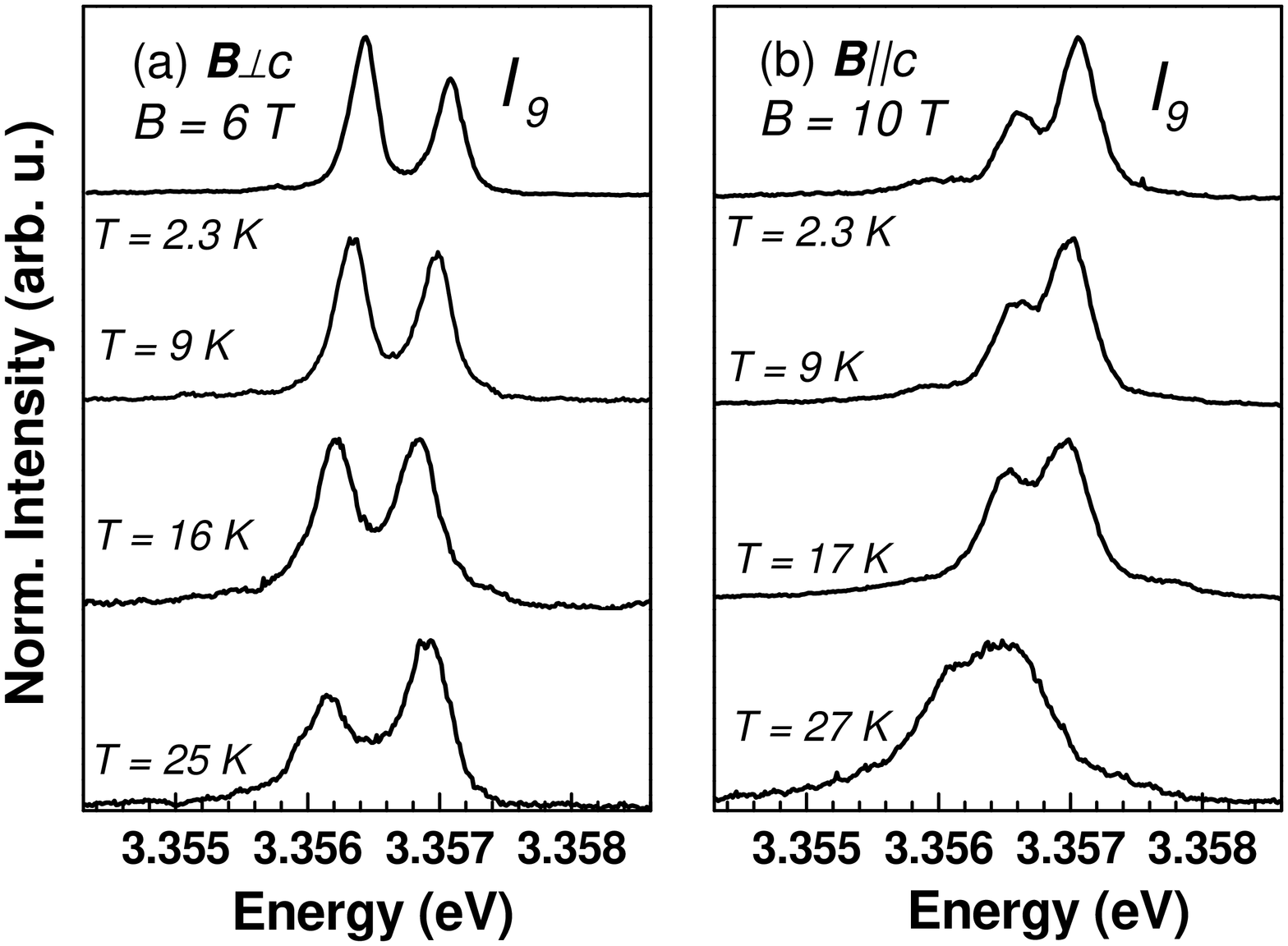}
\end{center}
\caption{\label{Fig9} Temperature-dependent  PL  spectra of the bound exciton line  $I_9$  in  ZnO:N
       for magnetic field $B = 6$ T perpendicular to the $c$ 
       axis of the crystal (a) and magnetic field  $B = 10$ T parallel 
to the $c$ axis of the
       crystal (b). Note that the spectral resolution in
       (a) and (b) is limited to $0.3$ meV.}
\end{figure*}

For ${\bm B} \perp c$, we have found that the  intensity  of  the low-energy  
PL 
component  of the $I_9$ line in ZnO:N  is  stronger than that of  
the  high-energy  component  (see Fig. \ref{Fig9} (a)). This was predicted  for  
an
acceptor bound exciton (see Figure \ref{Fig5} (c) and (d)). On  a  
first
glance,  this  contradicts  the  respective  results   of   the
transmission  experiments  presented  above,  but  is  in  good
agreement   with  Refs.  \onlinecite{1,2}. However, we did not observe 
this effect for the $I_9$ PL line in as grown ZnO  where the intensity 
ratio of both  component was independent of the temperature as well as 
of the magnetic field value. For ${\bm B} \parallel c$, the 
temperature behavior of  the
PL components of the $I_9$ for as grown ZnO and ZnO:N  is in an 
excellent agreement  with
the expected behavior of the $(D^0,X_A (\Gamma_7)) \longrightarrow  
D^0$ transitions with 
$|g_h^\parallel |<g_e$ (see  Fig.  \ref{Fig5} (a)).  The  low-energy
Zeeman  component exhibits less intensity in ${\bm B} \parallel c$ 
(see Fig. \ref{Fig9}(b)). Its intensity increases with increasing 
temperatures.  In contrast to that, 
 the  high-energy  PL component would to be expected  weaker  for  the 
$(A^0(\Gamma_7),X_A (\Gamma_7)) \longrightarrow  A^0(\Gamma_7)$ 
transitions (see Figure \ref{Fig5}(c)). But this was not observed. 
Therefore,  we conclude that the PL results 
for the $I_9$ line in ZnO:N for ${\bm B} \perp c$ and ${\bm B} 
\parallel c$ contradict each other. Taken alone they do not allow an 
unambiguous assignment   of the bound exciton complex. 

\begin{figure*}[hp]
\begin{center}
\vskip 1cm
\includegraphics*[width=5.5cm,height=7.5cm,angle=270]{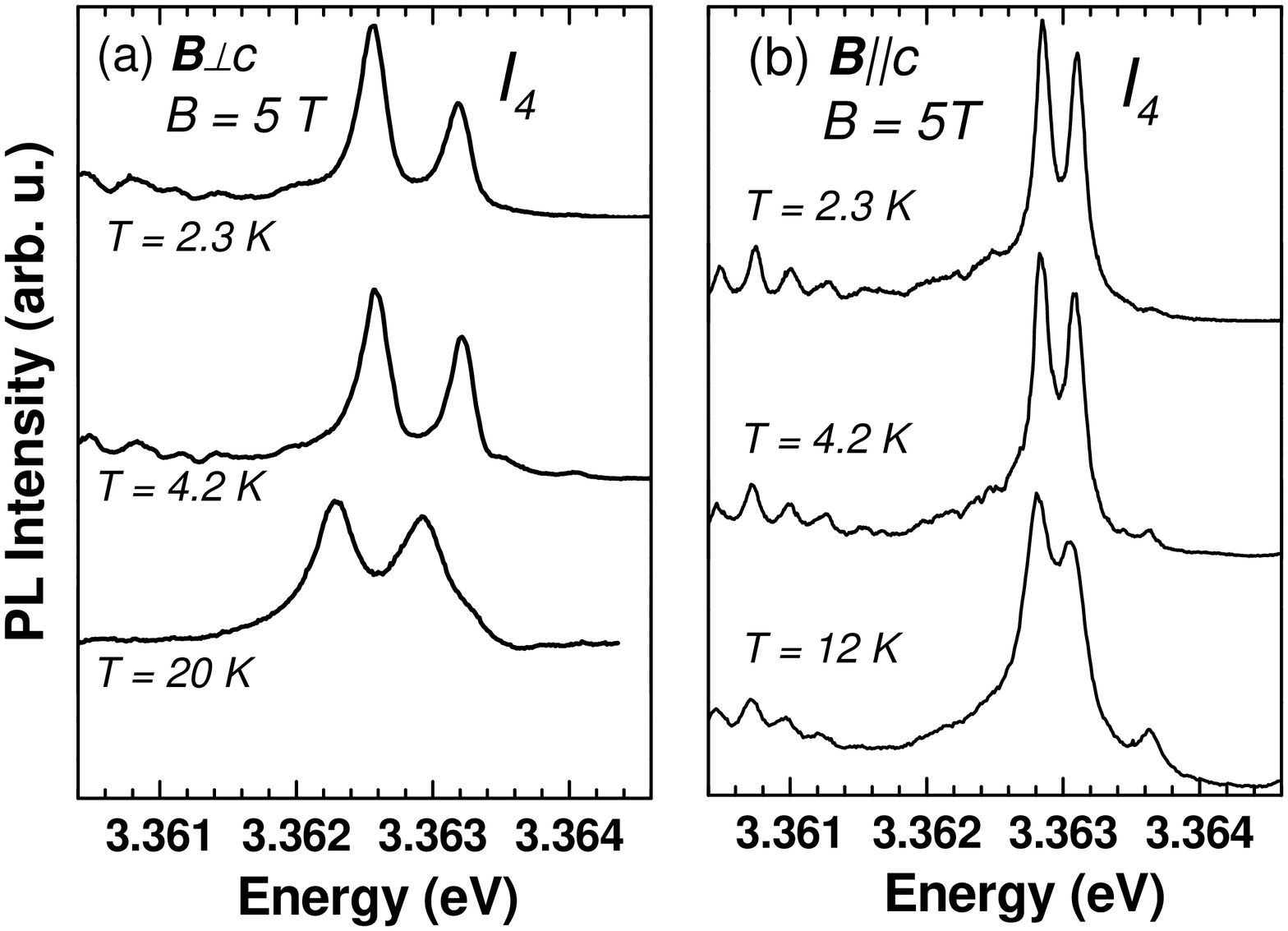}
\end{center}
\caption{\label{Fig10} Temperature-dependent  PL  spectra of the bound exciton line  $I_4$  in as grown ZnO
       for magnetic field $B = 5$ T perpendicular to the $c$ 
       axis of the crystal (a) and parallel 
to the $c$ axis of the
       crystal (b). }
\end{figure*}

We have detected a very similar  controversial  thermalization 
behavior of the $I_4$ PL line (see Figure  \ref{Fig10}). 
For ${\bm B} \perp c$  at low temperatures (Fig. \ref{Fig10} (a)),  a
stronger intensity was recorded for the low-energy Zeeman split
component  contradicting  to a donor-bound  
exciton  complex.
Although for ${\bm B} \parallel c$  (Fig. \ref{Fig10} (b)) the low-energy component was
stronger (as it is expected for an acceptor bound exciton), its 
intensity increased with increasing temperatures as it  would be typical 
for a  donor
bound  exciton.

 As the PL results  for the $I_9$ and $I_4$ lines are similar, we   
assign  both the $I_9$ and $I_4$ transition to the donor bound exciton 
complex on the basis of the self-consistent thermalization behavior of 
$I_9$ in transmission spectra.  
    An  explanation  of  the  controversial
temperature behavior of the PL components observed for 
 ${\bm B} \perp c$ can be, for example, attributed  to
the   light  re-absorption  process or to another energy transfer 
mechanism. It is necessary to note that the thermalization behavior 
corresponding to the $(D^0,X)$ complex  was not observed for the $I_4$ 
in  Ref. \onlinecite{3}. This was probably caused by the same 
processes that prevented our observation of the $I_4$ transition in PT 
spectra at all.

\section{\label{V} Discussion}

In this section we discuss our identification of the observed emission 
lines and compare  with  those
known  from the literature. We also
evaluate   possible
candidates for generating the nitrogen bound exciton transition  in
the   ZnO:N  crystals.  

\subsubsection*{$I_2$ and $I_3$ lines}

We attributed the $I_2$ and $I_3$ emission lines observed in ZnO:N to 
 transitions from the exchange split states of the 
$(D^+,X(\Gamma_7))$ complex. 
The formation of the ionized donor bound exciton complex is caused by 
the N doping in the same way as by the Na and Li doping in ZnO:Na and 
Zn:Li, respectively.  The  value of $0.98$ meV obtained for the zero 
field spin-exchange  is close to $0.9$ meV reported  in Ref. 
\onlinecite{4} for the same lines and to the theoretical calculations 
of Ref. \onlinecite{24}. However,  the  authors of  Ref.  
\onlinecite{4}  assumed  another
valence  band  ordering  with the $\Gamma_9$ symmetry  of the  top 
valence subband and thus assigned the $I_2$ and $I_3$ lines to the 
$(D^+,X(\Gamma_9))$ complex. The  field
dependence  of  the  transition energies in  ${\bm B} \parallel c$ and 
their angular dependence at $B = 4.5$ T were  fitted
in Ref. \onlinecite{4} by using $g_h^\parallel  = 1.5$ for the 
$\Gamma_9$
hole. This would correspond to $g_{exc}=g_h^\parallel-g_e  < 0$ and, 
as one can see from the scheme
in Fig. \ref{Fig4} (b), to the $\sigma^-$ polarization of the upper-energy Zeeman split component of the $I_2$ transition. This  
contradicts to our observations in Fig. \ref{Fig2} (a).

In Ref. \onlinecite{3},  the absorption  lines  $I_{c1}$
($3.368$ eV) and $I_{c2}$ ($3.367$ eV), corresponding to the emission 
lines $I_2$ and $I_3$, respectively, were assigned to the different 
$(D^+,X)$ complexes: $I_{c1}$ to the  $(D^+,X_A(\Gamma_7))$ complex 
while $I_{c2}$ to the  $(D^+,X_B(\Gamma_9))$ complex. The latter 
represents  the
excited  state of the exciton bound to the charged  donor and indeed  
could be observed in  absorption.
However, the energy separation between $I_{c1}$   and  $I_{c2}$
 absorption lines is close to the  $0.98$  meV  spin--exchange 
 splitting  of  the $X_A(\Gamma_7)$  exciton  obtained  in  the
present work and much smaller than the separation of the $A$  and  $B$
valence  subbands ($4.5$ meV \cite{helde,overview}). Therefore,  we
assume  that the absorption lines $I_{c1}$  and $I_{c2}$ in Ref. 
\onlinecite{3} originate from the  $\Gamma_5$ and $\Gamma_1/\Gamma_2$ 
states  of
the $(D^+,X_A(\Gamma_7))$ complex. The preferred polarization ${\bm E} 
\parallel c$  of the
$I_{c2}$  transition  reported in Ref. \onlinecite{3} agrees with  
this  assumption,
too.

\subsubsection*{$I_4$ line}

The $I_4$ is known to stem from an exciton bound
to a  shallow  donor.\cite{6,7,helde,overview} A strong evidence of 
the donor bound
exciton  complex origin of the $I_4$ line is  given by the observation 
of TES transitions.\cite{7,8,helde,overview} The binding energy of the 
donor is determined to be about $46$ meV.\cite{helde,overview} 
Annealing annihilates this shallow donor and thus neither $I_4$ line 
nor the respective TES transition can be observed in the annealed 
samples. 
 Correlated by temperature-dependent Hall effect 
measurements\cite{helde} and the magnetic resonance 
experiments\cite{deti} this particular donor was attributed to 
Hydrogen.\cite{helde,overview} Our zero field PL measurements for the 
as grown ZnO and for the ZnO:N, whose preparation conditions included 
an annealing process, confirmed this correlation\cite{martin} and thus 
the $(D^0,X)$ origin of the $I_4$ line. 

We did not detect the $I_4$ line in PT spectra and thus could not 
obtain a direct evidence for its donor bound exciton nature from our 
magneto-transmission studies. Nevertheless,  the observed Zeeman 
splitting of the $I_4$ PL  line is an excellent agreement with the 
model of the $(D^0,X(\Gamma_7))$ complex. The obtained value of the 
$\Gamma_7$ hole effective $g$ factor is typical for the holes in  
excitons bound to ionized or neutral donors in ZnO.  We note, for 
example, that in CdS the  $g$ values were found to differ 
significantly for the holes in excitons bound to  shallow donors 
from the holes in shallow acceptor states.\cite{21} The thermalization 
behavior of the $I_4$ PL line is very similar to those of the $I_9$ 
line whose $(D^0,X)$  origin was confirmed directly by the magneto-PT 
studies.

\subsubsection*{$I_9$ line}

   Our assignment of the $I_9$ line to the donor bound exciton complex 
is based on the thermalization properties in absorption  in an 
excellent agreement with Ref. \onlinecite{3}. Taking into account  the 
re-
absorption  process in the bulk crystals, our PL results as well as 
the PL results reported in Refs. \onlinecite{1,2} can be also 
explained within the $(D^0,X(\Gamma_7))$ model. Similar to the case of 
the $I_4$ line,  we additionally support 
the  allocation  of the $I_9$ line to the donor bound  exciton
 by  the  observation of the TES  transitions in ZnO:N. These 
transition appear at energy about  $50$  meV  below  the  $I_9$
line  in zero field PL spectra,\cite{martin} which  corresponds
to  the  donor  binding  energy  of  about  $63$  meV. Recently, the 
TES transitions for the $I_9$  line  was
likewise  observed in nominally undoped ZnO\cite{7} and in  ZnO:Na and 
ZnO:Li  
single crystals.\cite{overview}

Previously, the  $I_9$  line in ZnO:Na and ZnO:Li  crystals  was 
associated  with sodium  acceptors.\cite{25} However, this assignment 
is in conflict to the recent observation of TES 
transitions.\cite{overview} We have found in the present work that  
the intensity of the  
$I_9$  PL and absorption transitions increase significantly in ZnO:N 
in comparison with undoped bulk crystals. The  same  Zeeman  behavior  
and  the  identical  energy
positions  in  as grown ZnO and doped ZnO:N  crystals 
prove that this line has the same origin. One could then associate 
this line with excitons bound to a N  acceptor in ZnO:N.  However, our 
identification of this line as a donor bound exciton complex excludes 
such assignment.

It is worth to note, that
the $I_9$  line  was reported   not  to gain in intensity upon  the  
annealing
processes at temperatures up to $850^\circ$ C.\cite{helde,overview} 
However, secondary ion mass spectroscopy experiments showed that In 
was unintentionally introduced into the ZnO:N upon implantation and 
annealing at  $900^\circ$ C. 
 In diffusion experiments performed quite recently  showed that the 
$3.357$ eV $I_9$ line could be indeed caused by the indium 
donor.\cite{overview} 

\subsubsection*{Do we observe the nitrogen bound excitons in ZnO:N ?}
 
Previous  publications  have  shown   that
nitrogen  on an oxygen site acts as an acceptor with a  binding
energy of about $165$ meV.\cite{22,26,27} PL studies on nitrogen doped  
ZnO
layers  showed that nitrogen leads to a  at $3.359$  eV
within an accuracy of $1$ meV.\cite{28}
    In this spectral
range  we have detected  the  strongest  intensity   for the  $I_7$ 
line in ZnO:N. The $I_{6a}$, $I_7$ and  $I_8$  lines show similar  
Zeeman
behavior in  comparison  to $I_4$  and  $I_9$ that is typical for the 
donor bound exciton complexes. However, since the
energy  differences of the lines $I_{6a}$, $I_7$ and $I_8$ are very 
small,
a  comprehensive  evaluation of magnetic field  dependence  and
thermalization   properties  is  very  difficult   and   partly
prevented   by  the  linewidth  (FWHM)  of  the  Zeeman   split
components. 

    In  the  following, two other possible candidates  for  the
nitrogen bound exciton are introduced and discussed. First, 
the $I_{9a}$ absorption  peak  appeared on the low-energy  tail  of  
the  $I_9$
 in the PT spectra of  ZnO:N (see Figs. \ref{Fig1} and \ref{Fig8}).  
We
emphasize  that this peak was not observed in the  as  grown
undoped ZnO as well as in annealed undoped ZnO.\cite{martin}
 However, we do not have a clear evidence of  its  acceptor bound  
exciton origin. Its position close 
to the $I_9$ line, weak intensity and the large half width (FWHM) 
prevents  consequent conclusions  about   its
thermalization properties in transmission.

    The last potential  candidate  for  the  nitrogen bound exciton 
transition is a  new  weak  line  located 
 about $2$ meV above the $I_9$ transition in the PL spectra of ZnO:N  
(see the line labeled as $I_{9}^*$ in  Fig.  \ref{Fig1} (b)). This 
line was too weak both in PL and transmission to perform the
 magneto-optical studies. We note however, that more likely this line is to be 
attributed to an excited rotator state of the $I_9$ donor bound 
exciton complex. This is in agreement with the calculated position of 
such state in Ref. \onlinecite{overview} and the excitation 
measurements of Ref. \onlinecite{1}. In a similar way, the weak 
transition located about $1.1$ meV above $I_4$ in as grown ZnO (the 
line labeled as $I_{4}^*$ in Fig. \ref{Fig1} (a)) may be attributed  
to the excited rotator state of the $I_4$ donor bound exciton complex. 

\section{\label{VI} Conclusion}

    In  summary,  experimental and theoretical  magneto-optical
studies on undoped and nitrogen doped bulk ZnO facilitated  new
insights  of  the valence band ordering,  hole  $g$
values,  and  of the nature of shallow bound exciton complexes.
The  $\Gamma_7$  character  of  the  upper  valence  band  was  
derived
according to polarization properties and angular dependence
in  magnetic  field  of excitons bound to charged  and  neutral
impurity  centers. Hence, the valence band $A$, $B$ and $C$ pose  the
$\Gamma_7$,  $\Gamma_9$  and $\Gamma_7$ symmetry, respectively.
 The obtained  $\Gamma_7$   hole
effective  g  value in parallel field of $-1.24\pm  0.04$ is in a good  
agreement  with
theoretical calculations.\cite{20} We observed no PL transitions  
involving  the $\Gamma_9$
hole  states. 

 The  presence  of ionized donors confirmed the formation  of
acceptors  in ZnO:N.
However,  an unambiguously allocation of the nitrogen  impurity
center to one of the observed bound exciton lines was not possible. 
The  dominant recombination  lines, $I_9$ 
 in ZnO:N  and $I_4$ in as grown ZnO, were assigned to 
 donor bound excitons. 
    
    \begin{acknowledgments}
 The work of A.~V.~Rodina was  carried out  during the stay at the  
Institute
for Solid State Physics, TU Berlin, and supported  by the Deutsche 
Forschungsgemeinschaft (DFG).
\end{acknowledgments}

\newpage

\begin{thebibliography}{100}
\bibitem{note1} The notation for the hole g factor and the order of 
the hole
Zeeman split components is the same as in Ref. \onlinecite{20}
\bibitem{note2} This value can be calculated using the theory 
developed in Ref. \onlinecite{MMR} with effective mass Luttinger 
parameters from Ref. \onlinecite{20}.
\end{thebibliography}
\end{document}